\documentclass[a4paper,12pt]{article}

\PassOptionsToPackage{table}{xcolor}
\usepackage{jheppub} 

\usepackage[T1]{fontenc} 

\usepackage[all]{xy}
\usepackage[percent]{overpic}
\usepackage{slashed}
\usepackage{wrapfig}
\usepackage{tabu}
\usepackage{diagbox}
\usepackage{mathrsfs,amsmath,amssymb,amsthm,amsfonts,tikz,graphicx,accents,hyperref, color}
\usepackage{dsfont,epiolmec, latexsym, stmaryrd, comment}
\usepackage{slashed,ccaption}
\usepackage{mathrsfs, calligra}
\usepackage{leftidx}
\usepackage{import}
\usepackage{multirow}
\usepackage{amsfonts}
\usepackage{pifont}
\usepackage{tabularx}
\usepackage[utf8]{inputenc}
\usetikzlibrary{intersections,calc}
\usepackage{tikz-3dplot}
\usepackage{ifthen}
\usetikzlibrary{arrows}
\usepackage{amsmath}
\usepackage{xcolor}

\usepackage{caption}

\usepackage{array}
\usepackage[percent]{overpic}
\usepackage{wrapfig}
\usepackage{bbm}
\usepackage{tabu}
\usepackage{slashed}
\usepackage{fancyhdr} 
\usepackage{amsmath}
\usepackage{amsfonts}
\usepackage{amssymb}
\usepackage{diagbox}

\hypersetup{ linktoc=all,
    colorlinks, linkcolor={blue}, 
    citecolor={red}, urlcolor={darkpink}
}

\graphicspath{{Images/}}

\definecolor{light gray}{RGB}{220,220,220}
\definecolor{dark purple}{RGB}{108,0,217}
\definecolor{pink}{RGB}{190,20,100}
\definecolor{orang}{RGB}{193,63,0}
\definecolor{green}{RGB}{11,98,17}
\definecolor{darkpink}{RGB}{153,0,76}
\definecolor{bluegreen}{RGB}{0,102,102}
\definecolor{greenlagan}{RGB}{0,102,0}
\definecolor{redgreen}{RGB}{102,102,0}
\definecolor{Redgreen}{RGB}{153,76,0}
\definecolor{vividviolet}{rgb}{0.62, 0.0, 1.0}
\definecolor{amaranth}{rgb}{0.9, 0.17, 0.31}
\definecolor{palatinateblue}{rgb}{0.15, 0.23, 0.89}
\definecolor{brightpink}{rgb}{1.0, 0.0, 0.5}
\definecolor{cornflowerblue}{rgb}{0.39, 0.58, 0.93}
\definecolor{deepcarminepink}{rgb}{0.94, 0.19, 0.22}
\definecolor{radicalred}{rgb}{1.0, 0.21, 0.37}
\usepackage{graphicx}
\usepackage{tikz}
\usetikzlibrary{arrows,chains,shapes,matrix,positioning,scopes}
\usetikzlibrary{decorations.markings}
\tikzstyle arrowstyle=[scale=1]
\tikzstyle directed=[postaction={decorate,decoration={markings,
    mark=at position .65 with {\arrow[arrowstyle]{stealth}}}}]
\tikzstyle reverse directed=[postaction={decorate,decoration={markings,
    mark=at position .65 with {\arrowreversed[arrowstyle]{stealth};}}}]
\usepackage{import}
\usepackage{accents}
\usepackage{mathrsfs,amsmath,amssymb,slashed}
\usepackage{multirow,multicol}
\usepackage{enumitem}
\usepackage[percent]{overpic}
\usepackage{slashed}

\usepackage{wrapfig}
\usepackage{tabu}
\usepackage{diagbox}
\usepackage{comment}
\usepackage{tikz}
\usepackage{mathrsfs,amsmath,amssymb,amsthm,amsfonts,graphicx,accents,hyperref,color}
\usepackage{leftidx}
\usepackage{import}
\usetikzlibrary{decorations.pathmorphing}
\DeclareFontFamily{OT1}{rsfs}{}

\DeclareFontShape{OT1}{rsfs}{m}{n}{ <-7> rsfs5 <7-10> rsfs7 <10->rsfs10}{} 

\DeclareMathAlphabet{\mycal}{OT1}{rsfs}{m}{n}

\newcommand{\tcr}[1]{{\color{red}{#1}}}

\newcommand{\de}{\text{d}}

\newcommand{\eps}{\varepsilon}

\newcommand{\p}{\partial}

\newcommand{\be}{\begin{equation}}
\newcommand{\ee}{\end{equation}}

\newcommand{\cJ}{\mathcal{J}}
\newcommand{\cL}{{\mathcal L}}

\newcommand{\tcb}{\textcolor{blue}}

\newcommand{\df}{\mathbbmss{d}}
\newcommand{\di}{\mathbbmss{i}}

\newcommand\hnote[1]{\textcolor{blue}{\bf [H:\,#1]}}

\newcommand\vnote[1]{\textcolor{cyan}{\bf [V:\,#1]}}
\newcommand\snote[1]{\textcolor{darkpink}{\bf [Sh:\,#1]}}
\title{{\LARGE{Sliding Surface Charges on AdS$_3$ }}}
\author{H. Adami$^{a}$, V. Hosseinzadeh$^{a}$ and M. M. Sheikh-Jabbari$^{a,b}$}

\affiliation{\it $^a$ School of Physics, Institute for Research in Fundamental
Sciences (IPM),\\ P.O.Box 19395-5531, Tehran, Iran}

\affiliation{\it $^b$ The Abdus Salam ICTP, Strada Costiera 11, 34151, Trieste, Italy}

\emailAdd{hamed.adami@ipm.ir}
\emailAdd{v.hosseinzadeh@ipm.ir}
\emailAdd{jabbari@theory.ipm.ac.ir}

\preprint{IPM/P-2020/005}

\abstract{We consider 
Einstein gravity on a patch of AdS$_3$ spacetime between two radii $r_1, r_2$. We compute surface charges and their algebra at an arbitrary radius $r$ such that it reduces to a given set of surface charges at $r_1, r_2$. The $r$-dependent charges  become integrable upon addition of an appropriate boundary $Y$-term. We observe that soft excitations at each boundary are independent of those at the other boundary. We explicitly construct solution geometries which interpolate between these radii with specified surface charges. The interpolation is smooth provided that the mass and angular momentum measured at the two boundaries are equal. }
\begin{document}
\maketitle
\section{Introduction}

As we learn in the undergraduate electromagnetism course,  formulating Maxwell's  theory in a spacetime with boundaries requires imposing appropriate boundary conditions on the bulk fields and that these boundary conditions are outcome of degrees of freedom which reside at the boundary. The idea of ``asymptotic symmetries'' or ``soft charges'' provides us with a way to enumerate/label the \textit{boundary degrees of freedom} through the boundary conditions we impose, without knowing about the details of the boundary theory which has effectively given rise to the boundary conditions, see e.g. \cite{Barnich:2010bu, Barnich:2019qex, Barnich:2019xhd} for further discussions. 

In theories of gravity, motivated by quantifying the information carried to asymptotic null infinity through gravity waves, Bondi, van der Burg, Metzner and Sachs (BMS) \cite{Bondi:1962, Sachs:1962} have studied asymptotic symmetries and algebras for asymptotically flat metrics with certain falloff behavior. Similar analysis of asymptotic symmetries on asymptotically AdS$_3$ spacetimes leads to two Virasoro algebras at the Brown-Henneaux central charge \cite{Brown:1986nw}. 
As the electromagnetic examples alluded above indicate, in a gravitational theory one may also consider imposing appropriate boundary conditions elsewhere and not just at the asymptotic region of spacetime. In this case,  putting a material boundary (which has energy momentum tensor) would backreact on the bulk geometry as well. However, structure of the metric in special loci in spacetime may naturally lead to specific set of boundary conditions. Horizons are examples of such special loci. Presence of new ``near horizon degrees of freedom'' which are labeled by the near horizon symmetries, has been the main idea behind the soft hair proposal \cite{Hawking:2016msc} to describe black hole microstates and to address the information paradox. In stationary black holes we typically deal with a Killing horizon, a null surface of constant surface gravity generated by a Killing vector field. It is then natural to require that boundary conditions on metric fluctuations near a (Killing) horizon should respect the properties mentioned. This has been carried out in \cite{Afshar:2016wfy, Afshar:2016kjj, Grumiller:2019fmp} yielding a specific near horizon symmetry algebra. One may then try to analyze near horizon symmetries of more general class of horizons with less restrictive near horizon boundary conditions, as e.g. done in \cite{Donnay:2015abr, Donnay:2016ejv, Chandrasekaran:2018aop, Chandrasekaran:2019ewn, Donnay:2019zif, Donnay:2019jiz, Adami:2020amw}.

In addressing questions regarding black holes, we hence typically deal with two boundaries, one at the horizon and the other in the asymptotic region. Each of these boundary conditions keep different physical properties fixed at each boundary: Asymptotic boundary conditions like those of Brown-Henneaux \cite{Brown:1986nw} or BMS \cite{Bondi:1962, Sachs:1962, Barnich:2011mi} generically keep the ADM charges (mass and angular momentum) fixed while the near horizon ones keep the temperature (and horizon angular velocity) fixed \cite{Afshar:2016kjj, Afshar:2016uax, Sheikh-Jabbari:2016npa, Afshar:2017okz, Grumiller:2019fmp}. Each of these yield different set of charges and algebras at each boundary. The question of imposing two different boundary conditions at two different loci was recently addressed in \cite{Grumiller:2019ygj, Henneaux:2019sjx}, both of which are in the context of AdS$_3$ gravity and employ  Chern-Simons description of the theory. 

In this paper we address a similar problem of gravity on a patch of AdS$_3$ spacetime restricted between two arbitrary radii, as depicted in Fig. \ref{figr1r2}. In this sense our work is a continuation and extension of \cite{Grumiller:2019ygj, Henneaux:2019sjx}, but we put more emphasis on the metric formulation and the geometry. We impose boundary conditions which yield two chiral $U(1)$ current algebras at each boundary, much like the near horizon case analyzed in \cite{Afshar:2016wfy, Afshar:2016kjj}. We analyze surface charges at an arbitrary radius between the two boundaries. We show  smoothness of the geometry requires that the zero mode charges of these $U(1)$ current algebras associated with each boundary should be equal two each other. Moreover, while the charge algebra at the two boundaries are the same, the set of charges associated with one boundary, except for the zero modes, is not at all correlated with the charges at the other boundary. We demonstrate this by explicitly constructing the family of solutions to AdS$_3$ gravity which smoothly interpolate between the two boundaries of arbitrary charges. In the following sections we establish these results and in the last section discuss the physical meaning and implications of these for the soft hair proposal. 

\section{Review of \texorpdfstring{AdS$_3$}{AdS3} gravity and associated surface charges}\label{sec:2}

\begin{figure}[t]
    \centering
    \includegraphics[scale=.15]{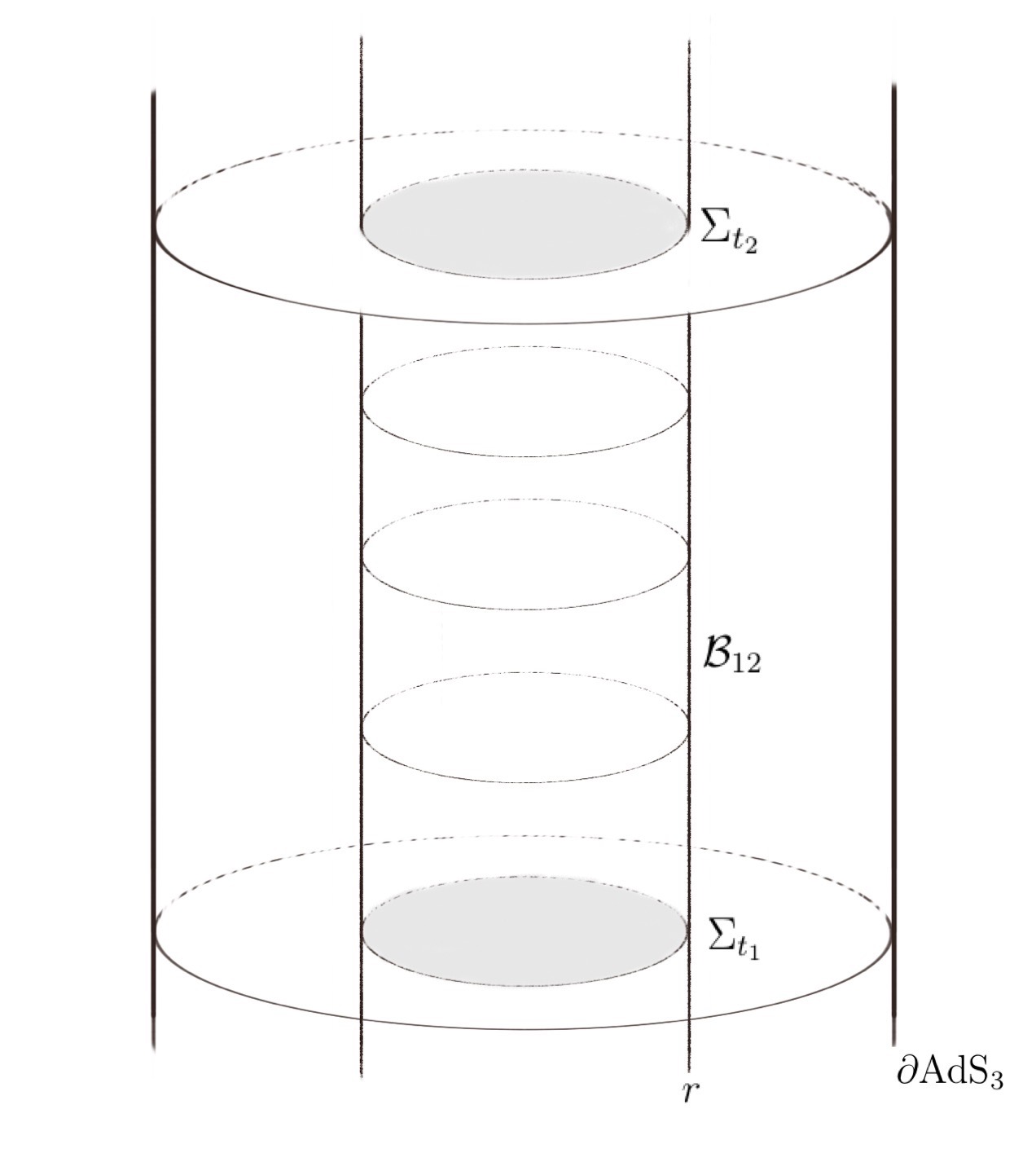}
    \caption{The horizontal  shaded surfaces $\Sigma_{t_1}, \Sigma_{t_2}$ show two constant time slices with  radii less than $r$. ${\cal B}_{12}$ is the gray cylinder at radius $r$ between $\Sigma_{t_1}, \Sigma_{t_2}$.}
    \label{fig-t1t2-r}
\end{figure}

The AdS$_3$ gravity is described by the action
\begin{equation}\label{3D-gravity-action}
    S = \int_\mathcal{M}  L[g] \de^3 x +S_{\text{bdy}}\, ,
\end{equation}
where
\begin{equation}\label{Lagrangian}
    L[g]= \frac{\sqrt{-g}}{16 \pi G} \left( R+2 \Lambda \right)\, ,
\end{equation}
is the Lagrangian in which $R$ is Ricci scalar and $\Lambda =-1/l^2$ is cosmological constant.  $S_{\text{bdy}}$ is boundary term which ensures a well-posed action principle, i.e. 
\begin{equation}\label{bdy-term}
    S_{\text{bdy}} =  \int_{\mathcal{B}} L^\mu_{\text{bdy}}[g] \de^2 x_{\mu} \, ,\qquad \text{with} \qquad L^\mu_{\text{bdy}}[g] = \frac{\sqrt{-g}}{16 \pi G }K \, n^{\mu}
\end{equation}
where ${\mathcal{B}}$ is the boundary of ${\mathcal{M}}$. Here we are assuming that topologically $\mathcal{M} = \mathbb{R} \times \Sigma$ where $\Sigma$ is a two-dimensional spacelike manifold. The boundary of this manifold $\partial \Sigma$ may consist of an $S^{1}$ or more circles. In this work we focus on formulating the boundary value problem on a patch of AdS$_3$ where the boundary consists of two coaxial cylinders of radii $r_1, r_2$, as depicted in Fig. \ref{figr1r2}. However, we develop the formulation of computing surface charges at a surface of arbitrary radius $r$, depicted in Fig. \ref{fig-t1t2-r}. $n^{\mu}$ is normal vector to the timelike boundary $\mathcal{B}= \mathbb{R} \times \partial\Sigma $ and $K= \nabla_{\mu}n^{\mu}$ is trace of extrinsic curvature of $\mathcal{B}$. We note that the boundary term \eqref{bdy-term} is one-half the Gibbons–Hawking–York boundary term \cite{Gibbons:1976ue}. The factor of one-half is needed to make the AdS$_3$ gravity action \eqref{3D-gravity-action}  regular \cite{Miskovic:2006tm}.

\subsection{Surface charge analysis, Lee-Wald symplectic form}\label{sec:2.1}

Variation of the total Lagrangian $L_{_{\text{\tiny{T}}}}[g]=L[g]+\partial_\mu L^\mu_{\text{bdy}}[g]$ is
\begin{equation}\label{Lagrangian-variation}
    \delta L[g] = -\frac{\sqrt{-g}}{16 \pi G} \mathcal{E}^{\mu \nu} \delta g_{\mu \nu} + \partial_{\mu} \Theta^{\mu} [g; \delta g]\, ,
\end{equation}
where
\begin{equation}\label{EOM}
    \mathcal{E}^{\mu \nu} = G^{\mu \nu} - \frac{1}{l^2} g^{\mu \nu}\, .
\end{equation}
Equations of motion are $\mathcal{E}^{\mu \nu}=0$. Eq.\eqref{Lagrangian-variation} defines the  `surface term' $\Theta^\mu$ only up to a total divergence,
\begin{equation}\label{Theta-Y}
   \Theta^{\mu} [g; \delta g]= \Theta^{\mu}_{_{\text{\tiny{LW}}}} [g; \delta g]+ \delta L^\mu_{\text{bdy}}[g]+ \partial_\nu Y^{\mu\nu} [g; \delta g],
\end{equation}
 where
\begin{equation}
    \Theta^{\mu}_{_{\text{\tiny{LW}}}} [g; \delta g]=\frac{\sqrt{-g}}{8 \pi G} \nabla^{[\alpha} \left( g^{\mu ] \beta} \delta g_{\alpha \beta} \right)\,,
\end{equation}
is the ``Lee-Wald symplectic potential'' \cite{Lee:1990nz}. As we see the above analysis does not specify $Y$-term (usually called boundary term). We will discuss in the next subsection that $Y$ may be fixed (restricted) through other physical  requirements.

The symplectic current
\begin{equation}\label{LW-Symp}
    \omega^{\mu }[g ; \delta_1 g , \delta_2 g]=\delta_1 \Theta^{\mu}[g;\delta_2 g] -\delta_2 \Theta^{\mu}[g;\delta_1 g]\, , 
\end{equation}
is a skew-symmetric function of metric variations and is conserved on-shell, i.e.
\begin{equation}\label{Cons-Curr}
    \partial_\mu \omega^{\mu }[g ; \delta_1 g ,\delta_2 g] \approx 0\, , 
\end{equation}
where $\approx$ denotes that the equality holds on-shell. That is, when $g_{\mu \nu}$ satisfies equation of motion $\mathcal{E}^{\mu \nu}=0$ and $\delta g_{\mu \nu}$'s satisfy linearized equations of motion $\delta \mathcal{E}^{\mu \nu}\approx 0$. Being a total variation, the $\delta L^\mu_{\text{bdy}}[g]$ term in \eqref{Theta-Y} does not contribute to $\omega$ \eqref{LW-Symp}.

Pre-symplectic form can be defined through integrating symplectic current over a spacelike co-dimension one surface $\Sigma$ with boundary $\partial \Sigma$, \footnote{The volume element is defined as $\de^{(D-p)} x_{\mu_1 \cdots \mu_p} = \frac{1}{p! (D-p)!} \varepsilon_{\mu_1 \cdots \mu_p 
\mu_{p+1} \cdots \mu_D} \de x^{\mu_{p+1}} \cdots \de x^{\mu_{D}} $ where $D$ and $p$ are dimensions of spacetime and integration surface respectively.}
\begin{equation}
   \Omega[g ; \delta_1 g ,\delta_2 g]= \int_{\Sigma}\omega^{\mu }[g ; \delta_1 g ,\delta_2 g] \de^2 x_{\mu}.
\end{equation}
The surface charge variation associated with the symmetry generator (diffeomorphism) $\xi$ is defined through pre-symplectic form as
\begin{equation}\label{charge-variation}
    \begin{split}
        \slashed{\delta} Q_{\xi} &:= \Omega[g ; \delta g ,\delta_\xi g]\\
        &= \oint_{\partial \Sigma} \left(\mathcal{Q}^{\mu \nu}_\xi[g ; \delta g]+ {\cal Y}^{\mu\nu}_\xi{[g ; \delta g]}\right) \de x_{\mu \nu}
    \end{split}
\end{equation}
with $\delta_\xi g_{\mu\nu}=\nabla_\mu \xi_\nu+\nabla_\nu\xi_\mu$. The first term is the contribution of the Lee-Wald part,
\begin{equation}\label{LW-part-charge}
    \mathcal{Q}^{\mu \nu}_\xi =\frac{\sqrt{-g}}{8 \pi G}\, \Big( h^{\lambda [ \mu} \nabla _{\lambda} \xi^{\nu]} - \xi^{\lambda} \nabla^{[\mu} h^{\nu]}_{\lambda} - \frac{1}{2} h \nabla ^{[\mu} \xi^{\nu]} + \xi^{[\mu} \nabla _{\lambda} h^{\nu] \lambda} - \xi^{[\mu} \nabla^{\nu]}h \Big),
\end{equation}
where $h_{\mu \nu}= \delta g_{\mu \nu}$  is a metric perturbation and $h= g^{\mu \nu}h_{\mu \nu}$, and the second term is
\begin{equation}\label{Y-term-charge}
{\cal Y}^{\mu\nu}_\xi\equiv \delta Y^{\mu \nu} [g; \delta_\xi g] - \delta_\xi Y^{\mu \nu} [g; \delta g] {-Y^{\mu \nu} [g; \delta_{\delta\xi} g]}. 
\end{equation}
{Here we are assuming a general \emph{field dependent variation}, that is when $\xi=\xi[g]$ {for which $\delta\xi\neq 0, \delta_{\delta\xi} g\neq 0$.}}
\subsection{Surface charge analysis, fixing the boundary Y-terms}\label{sec:2.2}

\begin{figure}[t]
    \centering
    \includegraphics[scale=.15]{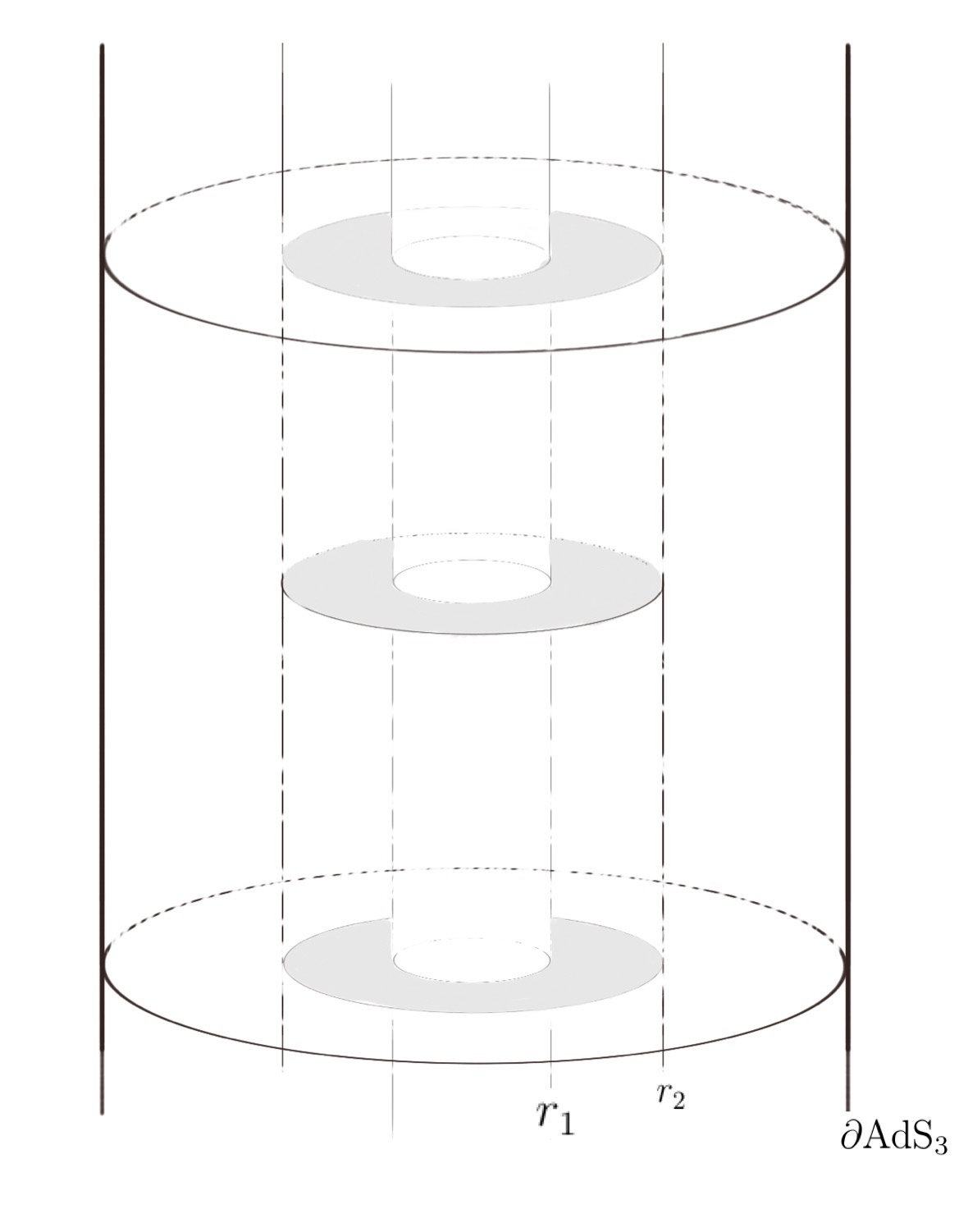}
    \caption{{Penrose diagram showing a patch of AdS$_3$ space between the two radii $r_1, r_2$. The vertical axis is the direction of time. We are computing the charges on a constant time slice at arbitrary $r$, $r_1<r<r_2$ assume having the charge ${\cal J}^\pm, {\cal K}^\pm$ respectively at $r_1, r_2$.}}
    \label{figr1r2}
\end{figure}

One may verify conservation of the variations $\slashed{\delta} Q_{\xi}$ and use this requirement to fix the $Y$-term. 
To this end, we compute charge variation at an arbitrary radius $r$ at two times, by integrating over two {different} constant time slices $\Sigma_{t_1}, \Sigma_{t_2}$, respectively $\slashed{\delta} Q^{(1)}_{\xi}, \slashed{\delta} Q^{(2)}_{\xi}$; see Fig.\ref{fig-t1t2-r}. Conservation means $\slashed{\delta} Q^{(1)}_{\xi}=\slashed{\delta} Q^{(2)}_{\xi}$. Starting from \eqref{charge-variation} and using \eqref{Cons-Curr}, we find
\begin{equation}\label{Conservation-equation}
\begin{split}
    \slashed{\delta} Q^{(1)}_{\xi}-\slashed{\delta} Q^{(2)}_{\xi} &\approx \int_{\mathcal{B}_{12}}\omega^{\mu }[g ; \delta g ,\delta_\xi g] \de^2 x_{\mu}\\
    &\approx \int_{\mathcal{B}_{12}}\left(\delta {\Theta}^{\mu}[g; \delta_\xi g] - \delta_\xi {\Theta}^{\mu}[g; \delta g]-{\Theta}^{\mu}[g; \delta_{\delta\xi} g] \right) \de^2 x_{\mu}
\end{split}
\end{equation}
where $\Sigma_{t_1} \cup \Sigma_{t_2} \cup \mathcal{B}_{12}$ are boundaries of a region of spacetime bounded between $\Sigma_{t_1}, \Sigma_{t_2}$ and $\mathcal{B}_{12}$ is time-like (or null) boundary of that region, \emph{e.g.} as depicted in Fig. \ref{fig-t1t2-r}. In the  `asymptotic symmetries' analysis as in the seminal  Brown and Henneaux work \cite{Brown:1986nw},  $\mathcal{B}_{12}$ is part of boundary of AdS$_3$ between the two constant time slices. In our current analysis ${\cal B}_{12}$ is at a generic $r$, as shown in Fig. \ref{fig-t1t2-r}, or may have two disconnected regions, as depicted in Fig. \ref{figr1r2}.  A sufficient condition for \eqref{Conservation-equation} to vanish, is
\begin{equation}\label{Conservation-condition}
    {\Theta} \cdot n \big|_{\mathcal{B}_{12}} \approx 0,
\end{equation}
where $n_\mu$ is the normal to $\mathcal{B}_{12}$.\footnote{{
To be more precise, vanishing of the right-hand side of \eqref{Conservation-equation} implies that projection of surface term on timelike boundary should be a total variation on the phase space, that is $\Theta[g ; \delta g] \cdot n = \delta \left( B[g] \cdot n \right)$. One can then replace the boundary term $L^\mu_{\text{bdy}}[g]$ by $L^\mu_{\text{bdy}}[g]-B^{\mu}[g]$ to have a well-defined variation principle. With this new boundary term, we again recover \eqref{Conservation-condition} as the  conservation condition.}} Since we know the expression for the Lee-Wald contribution, \eqref{Conservation-condition} may be used to specify the $Y$-term. We will use this in the next section in our charge analysis. 

Now let us examine the integrability of charges. The necessary and sufficient condition for integrability of the charge variation \eqref{charge-variation} is that the exterior derivative of the charge variation in the phase space is zero. By taking second variation of \eqref{charge-variation}, one finds \cite{Compere:2015knw}
\begin{equation}\label{integrability-condition}
{{\delta}_1 \slashed{\delta}_2 Q_{\xi} -{\delta}_2 \slashed{\delta}_1 Q_{\xi}  \approx - \oint_{\partial\Sigma} \,  2 \xi^{\mu}\omega^{\nu  }[g ; \delta_1 g ,\delta_2 g] \de x_{\mu \nu} + \slashed{\delta}_2 Q_{\delta_1\xi}-\slashed{\delta}_1 Q_{\delta_2\xi}}\, .
\end{equation}
Therefore the integrability condition implies that the right-hand side of the above equation should vanish. For the field independent cases {the two last terms} in \eqref{integrability-condition} drop out, recovering the Lee-Wald or Iyer-Wald integrability condition \cite{Lee:1990nz, Iyer:1994ys}.

\section{A general family of interpolating solutions}\label{sec:3}

We adapt radial coordinate $r$ so that $n_\mu \de x^{\mu} = \frac{l}{r}\de r$. Let $t$ and $\phi \sim \phi + 2 \pi$ denote time and angular coordinates respectively. All solutions to the field equations \eqref{EOM} are locally AdS$_3$ and we consider those which can be described through the following ansatz \cite{Afshar:2017okz}, 
\begin{equation}\label{M-01}
    ds^2 =  \frac{l^2}{r^2} dr^2 - \left( r \mathcal{A}^{+} - \frac{l^2 \mathcal{A}^{-}}{r}\right) \left( r  \mathcal{A}^{-} - \frac{l^2 \mathcal{A}^{+}}{r}\right) 
\end{equation}
where $\mathcal{A}^{\pm} = \lambda^{\pm} \de r + \zeta^{\pm}  \de t + \eta^{\pm} \de \phi$ are two one-forms. 
The Einstein field equations  then yield,
\begin{equation}\label{R-E-O-M}
    \de \mathcal{A}^{\pm}= 0 \, ,
\end{equation}

A simple choice of the coordinate system is when $\partial_r$ is hypersurface orthogonal, i.e. when $g_{rt}=g_{r\phi}=0$. For this choice, \eqref{R-E-O-M} may be solved as, $\eta^\pm=\eta^\pm (t,\phi),\ \zeta^\pm=\zeta^\pm (t,\phi)$ where,
\begin{equation}
    \partial_{t} \eta^{\pm} = \partial_{\phi} \zeta^{\pm} \qquad \text{with} \qquad \lambda^\pm =0.
\end{equation}
The above equations have infinitely many solutions, and the solution may be specified by other physical requirements. One such solution discussed in \cite{Afshar:2016wfy,Afshar:2016kjj} is $\zeta^\pm$ constant and fixed, and $\eta^\pm=J^\pm(\phi)$, with $J(\phi+2\pi)=J(\phi)$. This solution is appropriate for describing geometries with horizons at constant left and right temperatures $\zeta^\pm$. Another solution  is
\begin{equation}\label{A-special}
    \mathcal{A}^{\pm} =\eta^{\pm}(x^\pm) dx^\pm\,,\qquad \eta^\pm(x^\pm+2\pi)=\eta^\pm(x^\pm),
\end{equation}
where $x^\pm =t/l \pm \phi$. With this choice and for the special case of $\eta^\pm=J_0^\pm$, \eqref{M-01} describes a {Ba\~nados, Teitelboim and Zanelli (BTZ) black hole \cite{banados1992black}} of {outer} horizon radii $r_+= l$.\footnote{Note that in our coordinate system we only cover the outer horizon in $r^2>0$ region. The inner horizon is then located at $r^2=-l^2$. Note also that the radial coordinate $r$ in \eqref{M-01} and the usual BTZ radial coordinate are related as $r^2_{\text{\tiny{BTZ}}}=\frac{ (r^2J_0^-+J_0^+ l^2)(r^2J_0^++J_0^- l^2)}{r^2}$, see \cite{Sheikh-Jabbari:2014nya, Sheikh-Jabbari:2016unm}. Therefore, this solution corresponds to a BTZ black hole with surface gravity and horizon angular velocity,  $\kappa=\frac{2 J_0^-J_0^+}{l(J_0^++J_0^-)}  ,\ \Omega=\frac{J_0^+-J_0^-}{{l (J_0^++J_0^-)}}$.\label{footnote-BTZ}}

In this work, we are interested in solutions for which $\mathcal{A}_{\pm}$ reduces to \eqref{A-special} at two specific radii, say $r_1$ and $r_2$. To this end, we introduce
\begin{equation}\label{New-solution}
    \mathcal{A}^{\pm}= \mathcal{I}^{(f)}_\pm \de x^{\pm} + \mathcal{A}^{\pm}_r \de r
\end{equation}
where
\begin{equation}\label{cal-I}
 \mathcal{I}^{(f)}_\pm:=\mathcal{J}_{\pm}(x^\pm) f +\mathcal{K}_{\pm}(x^\pm) \left(1-f\right)
\end{equation}
and $f=f(r)$ is a smooth function of $r$ such that,
\begin{equation}\label{f-prop}
    f|_{r=r_1}=1, \quad f|_{r=r_2}=0 \,,\qquad
    \qquad f^{\prime}|_{r=r_1, r_2}=0.
\end{equation}
Here $\mathcal{J}_{\pm}$ and $\mathcal{K}_{\pm}$ are four periodic functions,
\begin{equation}\label{Periodicity-JK}
     {\cal J}_\pm (x^\pm+2\pi)={\cal J}_\pm (x^\pm),\qquad {\cal K}_\pm (x^\pm+2\pi)={\cal K}_\pm (x^\pm),
\end{equation}
which satisfy
\begin{equation}\label{Normalization}
    {\frac{1}{2 \pi}}\int_0^{2\pi} {\cal J}_\pm dx^\pm={\frac{1}{2 \pi}}\int_0^{2\pi} {\cal K}_\pm dx^\pm := J_0^\pm.
\end{equation}
The above and \eqref{cal-I} yield,
\begin{equation}\label{Normalization-I}
    {\frac{1}{2 \pi}}\int_0^{2\pi} {\cal I}_\pm(r) dx^\pm= J_0^\pm,
\end{equation}
which is independent of $f(r)$ and the radial coordinate $r$.

Equations of motion, or equivalently \eqref{R-E-O-M}, imply that the $r$-component of $\mathcal{A}^{\pm}$ should be of the form
\begin{equation}\label{New-solution-r-c}
    \mathcal{A}^{\pm}_r= \mathcal{Z}_{\pm}(x^\pm)  f^\prime \qquad \text{with} \qquad \partial_{\pm}\mathcal{Z}_{\pm}:= \mathcal{J}_{\pm} -\mathcal{K}_{\pm},
\end{equation}
where $f^\prime=\de f/\de r$. Therefore, $\mathcal{I}^{(f)}_\pm$ can be viewed as the \emph{interpolating} function. In the following, for the ease of notation we drop superscript $(f)$ on interpolating function. As one can see, for the special case of $f=const.$,  \eqref{New-solution} reduces to the solution \eqref{A-special}. Therefore, the set of geometries described by \eqref{New-solution} interpolates between two solutions of the form \eqref{A-special} with two different values of $\eta^\pm$, $\eta^\pm={\cal J}^\pm$ at $r_1$ and $\eta^\pm={\cal K}^\pm$ at $r_2$. The first equality in \eqref{Normalization}, that the zero mode of ${\cal J}, {\cal K}$ functions should be equal, is in fact the condition for having a smooth interpolating solution. This may also be seen from the definition of ${\cal Z}_\pm$ \eqref{New-solution-r-c}. As ${\cal Z}_\pm$ should also be periodic in $x^\pm$ and hence $\int_0^{2\pi} \partial_\pm {\cal Z}_\pm\  dx^\pm=0$. 

The metric components correspond to the solution \eqref{New-solution} with \eqref{New-solution-r-c} are
\begin{equation}\label{EH-01}
    \begin{split}
        g_{r r}= & \frac{l^2}{r^2} + (f^\prime)^2 \left[ l^2 \left( \mathcal{Z}_{+}^2 + \mathcal{Z}_{-}^2\right)+2 \mathcal{Z}_{+} \mathcal{Z}_{-} \mathcal{R}(r) \right]  \, , \\
        g_{r \pm} =& f^{\prime} \left[  l^2 \mathcal{Z}_{\pm} + \mathcal{Z}_{\mp} \mathcal{R}(r) \right]  \mathcal{I}_\pm\, , \\
        g_{\pm \pm} =& l^2  \mathcal{I}_\pm^2 \, , \\
        g_{+-}= & \mathcal{R}(r)  \mathcal{I}_- \mathcal{I}_+ \, ,
    \end{split}
\end{equation}
where
\begin{equation}
    \mathcal{R}(r)= - \frac{1}{2} \left( r^2 + \frac{l^4 }{r^2} \right).
\end{equation}

\section{Symmetries of the interpolating solutions}\label{sec:4}
In the previous section we introduced a four-function family of AdS$_3$ gravity solutions which interpolate between two, two-function family of solutions between radii $r_1, r_2$. In this section we show that there are two set of charges associated with this family. Consider diffeomorphisms, 
\begin{equation}\label{R.S.01}
    \xi^r=0, \qquad \xi^{\pm} = \pm\frac{T^{\pm}}{\mathcal{I}_\pm}\, ,
\end{equation}
with $\mathcal{I}_\pm$ given in \eqref{cal-I} and
\begin{equation}
    T^{\pm}=\epsilon_{\pm}  f+ \psi_{\pm}  (1-f) \, ,
\end{equation}
where $\epsilon_{\pm}$ and $\psi_{\pm}$ are arbitrary  periodic functions of $x^{\pm}$. These diffeomorphisms rotate us in the family of interpolating solutions \eqref{EH-01}, explicitly,  
\begin{equation}
    \delta_{\xi} g_{\mu \nu}=  g_{\mu \nu} (\mathcal{J}_{\pm}+\delta_{\xi}\mathcal{J}_{\pm}, \mathcal{K}_{\pm}+\delta_{\xi}\mathcal{K}_{\pm}) -g_{\mu \nu} (\mathcal{J}_{\pm}, \mathcal{K}_{\pm})
\end{equation}
with
\begin{equation}
       \delta_{\xi}\mathcal{J}_{\pm}= \pm\partial_{\pm}  \epsilon_{\pm} , \qquad \delta_{\xi}\mathcal{K}_{\pm}= \pm\partial_{\pm}  \psi_{\pm} \, .
\end{equation}
Therefore, 
\begin{subequations}
\begin{align}
 \delta_\xi\mathcal{I}_\pm=&\pm \partial_{\pm} T^\pm \, ,\\
    \delta_\xi\mathcal{Z}_{\pm}=&\pm \left( \epsilon_{\pm} -\psi_{\pm} \right).
\end{align}
\end{subequations}
That is, \eqref{R.S.01} generate symmetries in our interpolating family of solutions. 

\subsection{Algebra of symmetry generators}\label{sec:4.1}
The generators \eqref{R.S.01} are field dependent, they depend on background fields ${\cal J}^\pm, {\cal K}^\pm$ through {${\cal I}_\pm$}. Therefore, the algebra of the symmetry generators is given by the ``adjusted bracket''\cite{Compere:2015knw}:
\begin{equation}
    \left[ \xi_{1}, \xi_{2} \right]_{*} = \left[ \xi_{1}, \xi_{2} \right]_{\text{\tiny{Lie}}} - \delta^{(g)}_{\xi_1} \xi_2 +\delta^{(g)}_{\xi_2} \xi_1,
\end{equation}
where {$\left[\, , \, \right]_{\text{\tiny{Lie}}}$ is Lie bracket and} $\delta^{(g)}_{\xi_1} \xi_2$ denotes the change induced in $\xi_2$ due to the variation of metric $\delta_{\xi_1} g_{\mu \nu} = \mathcal{L}_{\xi_1} g_{\mu \nu}$. For the residual symmetry generators \eqref{R.S.01}, one can simply show that
\begin{equation}\label{xi-commute}
    \left[ \xi_{1}, \xi_{2} \right]_{*} =0\,.
\end{equation}
By expanding the vector field $\xi=\xi(\epsilon_\pm, \psi_\pm)$ in modes $X^{\pm}_{m}= \xi(e^{imx^\pm},0)$ and $Y^{\pm}_{m}= \xi(0,e^{imx^\pm})$, one obtains four copies of the $U(1)$ algebra
\begin{subequations}
    \begin{align}
       \left[ X^{\pm}_m, X^{\pm}_n \right]_{*} =0, && \left[ Y^{\pm}_m, Y^{\pm}_n \right]_{*} =0,\\
        \left[ X^{\pm}_m, Y^{\pm}_n \right]_{*} =0,  &&\left[ X^{\pm}_m, Y^{\mp}_n \right]_{*} =0,\\
        \left[ X^{+}_m, X^{-}_n \right]_{*} =0, && \left[ Y^{+}_m, Y^{-}_n \right]_{*} =0.
    \end{align}
\end{subequations}

\subsection{Isometries of the interpolating family}\label{sec:4.2}

The Killing vectors of the interpolating family $\zeta$ are a subset of \eqref{R.S.01} which keep the metric \eqref{EH-01} intact, $\mathcal{L}_{\zeta} g_{\mu \nu}=0$. That is, $\delta_{\zeta}\mathcal{J}_{\pm} =\delta_{\zeta}\mathcal{K}_{\pm}=0$. Moreover, since metric \eqref{EH-01} depends on $\mathcal{Z}_{\pm}$, we should also have $\delta_{\zeta} \mathcal{Z}_{\pm}= 0$. Therefore, 
\begin{equation}\label{EKV-01}
    \zeta^{r}= 0,\qquad \zeta^{\pm} =\frac{\alpha_{\pm}}{\mathcal{I}_{\pm}}\, ,
\end{equation}
where $\alpha_{\pm}$ are arbitrary constants. 

In general our family of solutions \eqref{EH-01} have two periodic (and hence globally defined) Killing vectors.\footnote{Note that all the geometries we consider here are locally AdS$_3$ and hence there are six local Killing vectors, generating $sl(2,\mathbb{R})\times sl(2,\mathbb{R})$ algebra. However, generically only two of these are also periodic in $x^\pm$ and are hence globally defined \cite{Banados:1992gq}. There is a similar situation for the class of Ba\~nados geometries \cite{Banados:1998gg} which is studied in detail in \cite{Sheikh-Jabbari:2014nya,Sheikh-Jabbari:2016unm}.} At generic values of radius $r$, one of these Killing vector fields is spacelike and the other is timelike. Requiring that for constant ${\cal J}_\pm, {\cal K}_\pm$ these Killing vectors reduce to the canonically normalized  Killing vectors of a BTZ black hole, fixes $\alpha_\pm=\kappa/2$ where $\kappa$ is the surface gravity. With this normalization, $ \zeta \cdot \zeta =- \frac{\kappa^2}{4r^2} \left( r^2 -l^2 \right)^2$. Therefore, we have a Killing horizon at $r_h=l$ (\emph{cf.}  footnote \ref{footnote-BTZ}).

\section{Charge analysis and \texorpdfstring{$Y$}{Y}-term}\label{sec:5}
Having specified the symmetry generators, we now compute the charges and their algebra for the set of solutions \eqref{EH-01}. 
The charge expression \eqref{charge-variation} on $\partial \Sigma $ can be written as
\begin{equation}\label{charge-01}
        \slashed{\delta} Q_{\xi} =  \int_{0}^{2 \pi} \left(\mathcal{Q}^{- r}_\xi+ {\cal Y}^{- r}_\xi\right) \de x^{+} -\int_{0}^{2 \pi} \left(\mathcal{Q}^{+ r}_\xi+ {\cal Y}^{+ r}_\xi\right) \de x^{-} \, .
\end{equation}
The above charge expressions are computed at a fixed, but arbitrary value of the radial coordinate $r$. 
Therefore only the two $\mathcal{Q}^{r\pm}_\xi$ components  contribute to the charge. A straightforward computation yields the following explicit form expressions for the interpolating solutions,
\begin{equation}\label{charge-densites}
\begin{split}
    \mathcal{Q}^{ \pm r} &= \mp\frac{l }{4 \pi G} T^{\mp} \delta \mathcal{I}_{\mp} - \frac{1 }{16 \pi G } \left[  \delta \sqrt{-g} \, \delta_\xi g^{\pm r} - \delta_\xi \sqrt{-g} \, \delta g^{\pm r} \right] \\
    &+\partial_{\mp} \left\{\frac{r^2 \mathcal{R}^\prime f^\prime }{32 \pi l G } \left[ \pm T^\pm \mathcal{I}_{\mp}\delta\left(\frac{ \mathcal{Z}_{\mp}}{\mathcal{I}_{\mp}} \right)\mp T^\mp \mathcal{I}_{\pm}\delta\left(\frac{ \mathcal{Z}_{\pm}}{\mathcal{I}_{\pm}} \right) \right] \right\}.
\end{split}
\end{equation}

The second line in \eqref{charge-densites} is a total derivative and hence do not contribute to the charge \eqref{charge-01}. The first term  in \eqref{charge-densites} yields an intergable charge, while the second term bars integrability. This second term is, however, of the form of a $Y$-term, \emph{cf.} \eqref{Y-term-charge}. So, fixing the $Y$-term as
\begin{equation}\label{Y-term}
    Y^{\mu \nu}[g ; \delta g]= \frac{\sqrt{-g}}{8 \pi G} \delta \hat{\epsilon}^{\mu \nu}
\end{equation}
where $\hat{\epsilon}^{\mu \nu} =n^{[\mu} m^{\nu]}$ with $ n_\nu dx^\nu=\frac{l}{r}dr,\ m=\frac{r}{l}\partial_r$, renders the charge integrable.
One may immediately verify that with the $Y$-term \eqref{Y-term}, the charge conservation condition \eqref{Conservation-condition} (vanishing flux through boundary) is also satisfied: {A direct computation leads to,
\begin{subequations}
\begin{align}
    \Theta_{_\text{\tiny{LW}}}^r[g;\delta g] =& -\frac{1}{16 \pi G } \delta \left( \sqrt{-g} K n^{r} \right)+ \frac{1}{16 \pi G  } \left[ \partial_+ \left( \sqrt{-g}\delta g^{+r} \right)+\partial_- \left( \sqrt{-g} \delta g^{-r} \right) \right]\, , \\
    \partial_\mu Y^{r \mu}= &-\frac{1}{16 \pi G  } \left[ \partial_+ \left( \sqrt{-g}\delta g^{+r} \right)+\partial_- \left( \sqrt{-g} \delta g^{-r} \right) \right]\, ,
    \end{align}
\end{subequations}
and hence $\Theta \cdot n = \frac{l}{r}\left( \Theta^{r}_{_{\text{\tiny{LW}}}}+ \delta L^r_{\text{bdy}}+ \partial_\mu Y^{r\mu} \right)=0$.} That is, the integrabilty \eqref{integrability-condition} and charge conservation equations are coming about through the same choice of the boundary $Y$-term. 

{A physically motivated requirement is that the $Y$-term \eqref{Y-term} should vanish at the boundaries. Since $Y$ is proportional to $f'$ this latter can be satisfied for generic values of $r_1,r_2$ if $f^{\prime}(r_1)=f'(r_2)=0$, as we have already required in \eqref{f-prop}.  By this assumption $\int_{\Sigma_t}\omega_{_{\text{\tiny{LW}}}}$ is conserved where $\partial\Sigma_t$ is the circle at $r=r_1$ or $r=r_2$ (see Fig. \ref{figr1r2}). In other words, the conserved and integrable charges of geometries \eqref{A-special} at $r=r_1$ and $r=r_2$ can be computed from $\int_{\Sigma_t}\omega_{_{\text{\tiny{LW}}}}$ and there is no need for the $Y$ term.}

Thus, charges associated with the symmetry generators \eqref{R.S.01} are
\begin{equation}\label{charge-f}
    Q_{\xi}= Q^+_\xi +Q^-_\xi \, ,
\end{equation}
with
\begin{equation}\label{Q-r}
    Q^\pm_\xi (r) = \frac{l }{4 \pi G}\int_0^{2 \pi} T^{\pm} \mathcal{I}_{\pm}\,\de x^\pm \, .
\end{equation}
These charges are integrable and conserved. We stress that the integrals \eqref{charge-01} are computed at a fixed but  arbitrary $r$ and hence the charges, as \eqref{Q-r} explicitly indicates, are $r$ dependent. In particular, 
\begin{equation}\label{Q-r1-r2}
 Q^\pm_\xi (r_1)=\frac{l }{4 \pi G}\int_0^{2 \pi} \epsilon^{\pm} \mathcal{J}_{\pm}\,\de x^\pm   \,, \qquad  Q^\pm_\xi (r_2)=\frac{l }{4 \pi G}\int_0^{2 \pi} \psi^{\pm} \mathcal{K}_{\pm}\,\de x^\pm   \,.
\end{equation}
The above expressions are precisely the charge expressions for geometries given by \eqref{A-special} \cite{Afshar:2017okz}.
That is, the charges \eqref{Q-r} are also interpolating between the charges associated with $f=0, f=1$ geometries.

\section{Charge algebra and redundancies of the phase space}\label{sec:6}

The algebra among charges can be obtained by the fundamental theorem of the covariant phase space method (see e.g. \cite{brown1986poisson, Compere:2018aar}) which states that 
\begin{equation}\label{charge-bracket-generic}
   \{ Q_{\xi_{1}}, Q_{\xi_{2}} \} = Q_{[\xi_{1},\xi_{2}]_*}+C(\xi_{1},\xi_{2})
\end{equation}
where the bracket is defined by $\delta_{\xi_{2}}Q_{\xi_{1}} = \{ Q_{\xi_{1}}, Q_{\xi_{2}} \}$ and $C(\xi_{1},\xi_{2})$ is a possible central extension. That is, the algebra among charge modes is isomorphic to the algebra of residual symmetry generators up to central extension terms. In our case the symmetry generators commute \eqref{xi-commute} and hence we only remain with the central extension, which is
\begin{equation}\label{charge-bracket-our-case}
\begin{split}
  \{ Q^\pm_{\xi_{1}}, Q^\pm_{\xi_{2}} \} &= \pm\frac{l }{4 \pi G}\int_0^{2 \pi} T_1^{\pm} \partial_\pm T_2^{\pm}\,\de x^\pm \, ,\\
      \{ Q^\pm_{\xi_{1}}, Q^\mp_{\xi_{2}} \} &=0.
\end{split}
\end{equation}

Upon expanding in modes $F^{\pm}_{m}:= Q^{\pm}(\epsilon^{\pm}=e^{i m x^{\pm}},\psi^\pm =0),\ H^{\pm}_{m}:= Q^{\pm}(\epsilon^{\pm}=0,\psi^\pm =e^{i m x^{\pm}})$, we find
\begin{equation}\label{F-H-I-01}
        F^{\pm}_{m}= \frac{l }{2 G}\, f\, \mathcal{I}^{\pm}_m, \qquad 
        H^{\pm}_{m}=  \frac{l }{2 G}\, (1-f)\, \mathcal{I}^{\pm}_m,
\end{equation}
where $\mathcal{I}^{\pm}_m= f \mathcal{J}^{\pm}_m + (1-f) \mathcal{K}^{\pm}_m$ and
\begin{equation}
    \mathcal{J}^{\pm}_m = \frac{1}{2 \pi} \int_0^{2\pi} e^{i m x^\pm} \mathcal{J}^{\pm} \de x^\pm \, , \qquad \mathcal{K}^{\pm}_m = \frac{1}{2 \pi} \int_0^{2\pi} e^{i m x^\pm} \mathcal{K}^{\pm} \de x^\pm\, .
\end{equation}
The charge algebra \eqref{charge-bracket-our-case} then reduces to
\begin{subequations}
    \begin{align}
       \{ F^{\pm}_{m}, F^{\pm}_{n} \} = & \mp \left(\frac{ i  l m}{2 G} \right) f^2 \, \delta_{m+n,0} \\
       \{ H^{\pm}_{m}, H^{\pm}_{n} \} = & \mp \left( \frac{ i  l m}{2 G}\right)  (1-f)^2 \, \delta_{m+n,0}\\
       \{ F^{\pm}_{m}, H^{\pm}_{n} \} =& \mp \left( \frac{ i  l m}{2G}\right) f (1-f) \, \delta_{m+n,0}
    \end{align}
\end{subequations}
and the brackets not displayed vanish.
By replacing the above brackets with commutators, $i\{\, ,\, \} \rightarrow [\, , \,]$, we obtain four $U(1)$ current algebras
\begin{subequations}\label{charge-algebra-F-H}
    \begin{align}
       [ F^{\pm}_{m}, F^{\pm}_{n} ] = &\pm m\, k f^2 \, \delta_{m+n,0} \\
       [ H^{\pm}_{m}, H^{\pm}_{n} ] = & \pm  m \, k  (1-f)^2 \, \delta_{m+n,0}\\
       [ F^{\pm}_{m}, H^{\pm}_{n} ] =& \pm m \, k f (1-f) \, \delta_{m+n,0}
    \end{align}
\end{subequations}
where $k= \frac{l }{2 G}=\frac{c}{3}$ and $c$ is the Brown-Henneaux central charge.\footnote{Note that the normalization of our charges is different than those used in \cite{Afshar:2016wfy} by a factor of 1/2, yielding a factor of 1/4 difference in commutators, in factor $k$.}

\paragraph{Redundancy in phase space.}  The family of interpolating solutions are specified by four functions ${\cal J}_\pm, {\cal K}_\pm$. At first sight this matches with the number of charges, $F^\pm, H^\pm$. However, we note that in metric \eqref{M-01}, and also in the charge \eqref{Q-r}, at any given $r$ only a certain combination of these four functions, namely ${\cal I}_\pm$, appear. This suggests that there is a redundancy in our description of the phase space and also the charges. To establish this, let us define,
\begin{subequations}\label{I-charge-01}
    \begin{align}
        I^{\pm}_{m} :=& F^{\pm}_m +H^{\pm}_m \, , \\
        D^{\pm}_{m} :=& (1-f) F^{\pm}_m -f H^{\pm}_m \, .
    \end{align}
\end{subequations}
Then, \eqref{F-H-I-01} yields,
\begin{equation}
    I^{\pm}_{m}= \frac{l }{2 G}\, \mathcal{I}^{\pm}_m\, , \qquad D^{\pm}_{m}=0\, .
\end{equation}
As we see the $D^\pm$ charges vanish on the phase space and the phase space points are governed by just two $r$ dependent functions ${\cal I}_\pm$.

\paragraph{Black hole entropy.} We discussed that our family of interpolating geometries in general correspond to a BTZ black hole with different kinds of soft hair excitations. These soft hair excitations can change as we move along the radial $r$ direction, nonetheless the value of the zero modes remain the same. 
In our coordinates, the horizon is at $r=l$ and line element at the bifurcation circle (a constant $t$ slice at $r=l$) is $ds^2{_{h}}= l^2({\cal I}_++ {\cal I}_-)^2d\phi^2$. Therefore, the Bekenstein-Hawking entropy is 
\begin{equation}\label{BH-entropy}
    S_{_\text{BH}}=\frac{l}{4G}\int_0^{2\pi} d\phi\ ({\cal I}_++{\cal I}_-)=\frac{2\pi l}{4G} (J_0^++J_0^-)={\pi} (I^{+}_0+I^{-}_0).
\end{equation}
The mass and angular momentum of the BTZ black hole are
\begin{equation}\label{BH-M-J}
    M=\frac{l}{4G} \left((J_0^+)^2+(J_0^-)^2\right),\qquad J=\frac{l}{4G} \left((J_0^+)^2-(J_0^-)^2\right).
\end{equation}
These correspond to exact symmetries (Killing vector) charges and hence their values are independent of the radius \cite{Hajian:2015xlp}, as expected.

\section{Discussion and outlook}

In this paper we continued the analysis of \cite{Grumiller:2019ygj, Henneaux:2019sjx} and studied AdS$_3$ gravity on a portion of AdS$_3$ space between two arbitrary constant $r$ slices $r_1, r_2$, as depicted in Fig. \ref{figr1r2}. In our family of solutions \eqref{M-01} with \eqref{New-solution}, constant ${\cal J}^\pm, {\cal K}^\pm$ case describes a BTZ black hole whose horizon is at $r_h=l$. For ${\cal J}^\pm={\cal K}^\pm$ case our analysis reduces to those of \cite{Afshar:2016wfy, Afshar:2016kjj} where the near horizon symmetries of BTZ black holes were studied. To see the connection to the near horizon symmetries,  
we note that one could have taken our inner boundary to be at the horizon, $r_1=l$, where our boundary conditions  reduce to those in \cite{Afshar:2016wfy, Afshar:2016kjj}. 

As emphasized  smoothness of geometries in the family of our solutions requires ${\cal J}^\pm$ and ${\cal K}^\pm$ to have equal zero modes \eqref{Normalization}, but are otherwise arbitrary and independent. In the Chern-Simons analysis of \cite{Grumiller:2019ygj, Henneaux:2019sjx} this condition arises from matching of holonomies around noncontractable circles in the (Euclidean) interpolating geometry. Algebraically, each geometry in our family may be viewed as an element in a coadjoint orbit of our $U(1)$ currents symmetry algebra, where each orbit is specified by the $J_0^\pm$ values. (Note that in the algebra \eqref{charge-algebra-F-H} $F^\pm_0, H^\pm_0$ commute with all elements of the algebra and hence these zero modes can be used to label the coadjoint orbits.) Therefore, our analysis reveals that as we move in the radial direction we may move in a given coadjoint orbit but it is not possible to smoothly traverse  the orbits. 

If we take $r_1=l$ while taking the other boundary to the AdS boundary, $r_2\rightarrow \infty$, we get a solution which covers the region outside the outer horizon of a BTZ black hole. In this case one may observe that we end up with a metric which  does not satisfy the Brown-Henneaux boundary conditions, see \cite{Afshar:2017okz} for more discussions. In this solution the near horizon soft hair are given by ${\cal J}^\pm$ functions while the asymptotic charges are given by ${\cal K}^\pm$, which are not constrained at all by the near horizon charges. 
The fact that near horizon and asymptotic soft charges are not correlated with each other was also observed and briefly discussed in \cite{Grumiller:2019ygj}. Here we find a similar phenomenon for an example with different boundary conditions at the two boundaries. 

Given the line of analysis and arguments yielding the independence of near horizon and asymptotic charges, we expect that this feature is not limited to the cases in AdS$_3$ gravity and should hold more generically. In particular, we expect this to be true for asymptotic flat 4d Kerr black hole, the near horizon symmetries of which was analyzed in \cite{Donnay:2015abr, Grumiller:2019fmp, Adami:2020amw} and its asymptotic symmetries are BMS$_4$ \cite{Barnich:2011mi}. This has implications for the soft hair proposal \cite{Hawking:2016msc}, according which the black hole microstates, the soft hair, are labeled by the near horizon soft charges, while the information carried by the Hawking radiation may be labeled by the asymptotic BMS$_4$ charges.

At a technical level in our charge computations we found that the charge variation computed from the Lee-Wald contribution is not integrable and conserved. This was a result of the fact that  the Lee-Wald flux does not vanish at the two boundaries at arbitrary $r_1, r_2$. This was remedied by the addition of a boundary $Y$-term \eqref{Y-term} which vanishes at two radii. For $r_1=l$  $r_2\to\infty$ this matches with the earlier familiar results in the literature. 

Finally, we would like to comment on the possibility of extension of asymptotic (boundary) symmetries into the bulk. To explain the point we start with an example. Let us consider the Virasoro excitations at the causal boundary of AdS$_3$ which are defined by the Brown-Henneaux boundary conditions \cite{Brown:1986nw}. One may then ask if it is possible to extend the Brown-Henneaux excitations to the bulk of AdS$_3$. The answer is of course affirmative and Ba\~nados geometries \cite{Banados:1998gg} achieve this. These are the family of AdS$_3$ solutions uniquely specified by their Virasoro charges. Moreover, as discussed in \cite{Compere:2015knw}, given the set of Ba\~nados geometries one may compute the associated Virasoro charge at an arbitrary radius, i.e. we are dealing with \emph{symplectic symmetries/charges} (and not just asymptotic symmetries/charges). The near horizon symmetries of \cite{Afshar:2016wfy, Afshar:2016kjj} are also symplectic in the same sense. Similar extension of the asymptotic symmetries to the bulk has also been studied in 4d examples \cite{Compere:2016hzt,Compere:2016jwb} and for near horizon extremal geometries \cite{Compere:2015bca, Compere:2015mza}. Given our analysis here and results of \cite{Grumiller:2019ygj, Henneaux:2019sjx}, one would expect that this should be a very general feature: (i) All the near horizon and/or asymptotic symmetries can be extended into the bulk and be made symplectic. (ii) This extension into the bulk is not unique at all. One can arbitrarily rotate in a given coadjoint orbit as we move in the radial direction. There is, of course, a specific extension, associated with the symplectic symmetries, in which one does not move in the orbit as we move in the radial direction. Explicitly, if we denote our radius dependent charges by ${\cal Q}(r)$, then ${\cal Q}(r_1), {\cal Q}(r_2)$ can generically be different elements in the same coadjoint orbit of the charge algebra. (iii) In the example we studied here, while the charges $I^\pm_n$ \eqref{I-charge-01} are $r$ dependent, the algebra of charges are $r$ independent. One can envisage examples where the algebra of charges are also $r$-dependent. The change in the algebra should, however,   be such that it does not force us traverse across the orbits of the algebra at either of the boundaries. A first example of the latter was given in \cite{Grumiller:2019ygj}.  It would of course be desirable to provide a rigorous proof for these statements in a general setting, beyond the examples so far studied in the literature.

\section*{Acknowledgement}
We are grateful to Hamid Afshar, Daniel Grumiller, Marc Henneaux, Raphaela Wutte and Hossein Yavartanoo for useful discussions and Ali Seraj for this contribution at the early stages of this work. MMShJ would like to thank the hospitality of ICTP HECAP and ICTP EAIFR where this research carried out. We acknowledge the support by 
INSF grant No 950124 and Saramadan grant No. ISEF/M/98204.
MMShJ acknowledge the Iran-Austria IMPULSE project grant, supported and run by Khawrizmi University.


\begin{thebibliography}{10}

\bibitem{Barnich:2010bu}
G.~Barnich, ``{The Coulomb solution as a coherent state of unphysical
  photons},'' {\em Gen. Rel. Grav.} {\bf 43} (2011) 2527--2530,
\href{http://www.arXiv.org/abs/1001.1387}{{\tt 1001.1387}}.

\bibitem{Barnich:2019qex}
G.~Barnich and M.~Bonte, ``{Soft degrees of freedom, Gibbons-Hawking
  contribution and entropy from Casimir effect},''
\href{http://www.arXiv.org/abs/1912.12698}{{\tt 1912.12698}}.

\bibitem{Barnich:2019xhd}
G.~Barnich, ``{Black hole entropy from nonproper gauge degrees of freedom: The
  charged vacuum capacitor},'' {\em Phys. Rev.} {\bf D99} (2019), no.~2,
  026007,
\href{http://www.arXiv.org/abs/1806.00549}{{\tt 1806.00549}}.

\bibitem{Bondi:1962}
H.~Bondi, M.~van~der Burg, and A.~Metzner, ``Gravitational waves in general
  relativity {VII.} {W}aves from axi-symmetric isolated systems,'' {\em Proc.
  Roy. Soc. London} {\bf A269} (1962) 21--51.

\bibitem{Sachs:1962}
R.~Sachs, ``Asymptotic symmetries in gravitational theory,'' {\em Phys. Rev.}
  {\bf 128} (1962) 2851--2864.

\bibitem{Brown:1986nw}
J.~D. Brown and M.~Henneaux, ``{Central Charges in the Canonical Realization of
  Asymptotic Symmetries: An Example from Three-Dimensional Gravity},'' {\em
  Commun. Math. Phys.} {\bf 104} (1986)
207--226.

\bibitem{Hawking:2016msc}
S.~W. Hawking, M.~J. Perry, and A.~Strominger, ``{Soft Hair on Black Holes},''
  {\em Phys. Rev. Lett.} {\bf 116} (2016), no.~23, 231301,
\href{http://www.arXiv.org/abs/1601.00921}{{\tt 1601.00921}}.

\bibitem{Afshar:2016wfy}
H.~Afshar, S.~Detournay, D.~Grumiller, W.~Merbis, A.~Perez, D.~Tempo, and
  R.~Troncoso, ``{Soft Heisenberg hair on black holes in three dimensions},''
  {\em Phys. Rev.} {\bf D93} (2016), no.~10, 101503,
\href{http://www.arXiv.org/abs/1603.04824}{{\tt 1603.04824}}.

\bibitem{Afshar:2016kjj}
H.~Afshar, D.~Grumiller, W.~Merbis, A.~Perez, D.~Tempo, and R.~Troncoso,
  ``{Soft hairy horizons in three spacetime dimensions},'' {\em Phys. Rev.}
  {\bf D95} (2017), no.~10, 106005,
\href{http://www.arXiv.org/abs/1611.09783}{{\tt 1611.09783}}.

\bibitem{Grumiller:2019fmp}
D.~Grumiller, A.~Pérez, M.~M. Sheikh-Jabbari, R.~Troncoso, and C.~Zwikel,
  ``{Spacetime structure near generic horizons and soft hair},'' {\em Phys.
  Rev. Lett.} {\bf 124} (2020) 041601,
\href{http://www.arXiv.org/abs/1908.09833}{{\tt 1908.09833}}.

\bibitem{Donnay:2015abr}
L.~Donnay, G.~Giribet, H.~A. Gonz{\'a}lez, and M.~Pino, ``{Supertranslations
  and Superrotations at the Black Hole Horizon},'' {\em Phys. Rev. Lett.} {\bf
  116} (2016), no.~9, 091101,
\href{http://www.arXiv.org/abs/1511.08687}{{\tt 1511.08687}}.

\bibitem{Donnay:2016ejv}
L.~Donnay, G.~Giribet, H.~A. Gonz{\'a}lez, and M.~Pino, ``{Extended Symmetries
  at the Black Hole Horizon},'' {\em JHEP} {\bf 09} (2016) 100,
\href{http://www.arXiv.org/abs/1607.05703}{{\tt 1607.05703}}.

\bibitem{Chandrasekaran:2018aop}
V.~Chandrasekaran, {\'E}.~{\'E}. Flanagan, and K.~Prabhu, ``{Symmetries and
  charges of general relativity at null boundaries},'' {\em JHEP} {\bf 11}
  (2018) 125,
\href{http://www.arXiv.org/abs/1807.11499}{{\tt 1807.11499}}.

\bibitem{Chandrasekaran:2019ewn}
V.~Chandrasekaran and K.~Prabhu, ``{Symmetries, charges and conservation laws
  at causal diamonds in general relativity},'' {\em JHEP} {\bf 10} (2019) 229,
\href{http://www.arXiv.org/abs/1908.00017}{{\tt 1908.00017}}.

\bibitem{Donnay:2019zif}
L.~Donnay and G.~Giribet, ``{Cosmological horizons, Noether charges and
  entropy},'' {\em Class. Quant. Grav.} {\bf 36} (2019), no.~16, 165005,
\href{http://www.arXiv.org/abs/1903.09271}{{\tt 1903.09271}}.

\bibitem{Donnay:2019jiz}
L.~Donnay and C.~Marteau, ``{Carrollian Physics at the Black Hole Horizon},''
  {\em Class. Quant. Grav.} {\bf 36} (2019), no.~16, 165002,
\href{http://www.arXiv.org/abs/1903.09654}{{\tt 1903.09654}}.

\bibitem{Adami:2020amw}
H.~Adami, D.~Grumiller, S.~Sadeghian, M.~Sheikh-Jabbari, and C.~Zwikel,
  ``T-witts from the horizon,'' \href{http://www.arXiv.org/abs/2002.08346}{{\tt
  2002.08346}}.

\bibitem{Barnich:2011mi}
G.~Barnich and C.~Troessaert, ``{BMS charge algebra},'' {\em JHEP} {\bf 1112}
  (2011) 105,
\href{http://www.arXiv.org/abs/1106.0213}{{\tt 1106.0213}}.

\bibitem{Afshar:2016uax}
H.~Afshar, D.~Grumiller, and M.~M. Sheikh-Jabbari, ``{Near horizon soft hair as
  microstates of three dimensional black holes},'' {\em Phys. Rev.} {\bf D96}
  (2017), no.~8, 084032,
\href{http://www.arXiv.org/abs/1607.00009}{{\tt 1607.00009}}.

\bibitem{Sheikh-Jabbari:2016npa}
M.~M. Sheikh-Jabbari and H.~Yavartanoo, ``{Horizon Fluffs: Near Horizon Soft
  Hairs as Microstates of Generic AdS3 Black Holes},'' {\em Phys. Rev.} {\bf
  D95} (2017), no.~4, 044007,
\href{http://www.arXiv.org/abs/1608.01293}{{\tt 1608.01293}}.

\bibitem{Afshar:2017okz}
H.~Afshar, D.~Grumiller, M.~M. Sheikh-Jabbari, and H.~Yavartanoo, ``{Horizon
  fluff, semi-classical black hole microstates --- Log-corrections to BTZ
  entropy and black hole/particle correspondence},'' {\em JHEP} {\bf 08} (2017)
  087,
\href{http://www.arXiv.org/abs/1705.06257}{{\tt 1705.06257}}.

\bibitem{Grumiller:2019ygj}
D.~Grumiller, M.~M. Sheikh-Jabbari, C.~Troessaert, and R.~Wutte,
  ``{Interpolating Between Asymptotic and Near Horizon Symmetries},''
\href{http://www.arXiv.org/abs/1911.04503}{{\tt 1911.04503}}.

\bibitem{Henneaux:2019sjx}
M.~Henneaux, W.~Merbis, and A.~Ranjbar, ``{Asymptotic dynamics of AdS$_3$
  gravity with two asymptotic regions},''
\href{http://www.arXiv.org/abs/1912.09465}{{\tt 1912.09465}}.

\bibitem{Gibbons:1976ue}
G.~W. Gibbons and S.~W. Hawking, ``Action integrals and partition functions in
  quantum gravity,'' {\em Phys. Rev.} {\bf D15} (1977)
2752--2756.

\bibitem{Miskovic:2006tm}
O.~Miskovic and R.~Olea, ``{On boundary conditions in three-dimensional AdS
  gravity},'' {\em Phys. Lett.} {\bf B640} (2006) 101--107,
\href{http://www.arXiv.org/abs/hep-th/0603092}{{\tt hep-th/0603092}}.

\bibitem{Lee:1990nz}
J.~Lee and R.~M. Wald, ``{Local symmetries and constraints},'' {\em J. Math.
  Phys.} {\bf 31} (1990)
725--743.

\bibitem{Compere:2015knw}
G.~Comp{\`e}re, P.-J. Mao, A.~Seraj, and M.~M. Sheikh-Jabbari, ``{Symplectic
  and Killing symmetries of AdS$_{3}$ gravity: holographic vs boundary
  gravitons},'' {\em JHEP} {\bf 01} (2016) 080,
\href{http://www.arXiv.org/abs/1511.06079}{{\tt 1511.06079}}.

\bibitem{Iyer:1994ys}
V.~Iyer and R.~M. Wald, ``Some properties of {N}{\"o}ther charge and a proposal
  for dynamical black hole entropy,'' {\em Phys. Rev.} {\bf D50} (1994)
  846--864,
\href{http://arXiv.org/abs/gr-qc/9403028}{{\tt gr-qc/9403028}}.

\bibitem{banados1992black}
M.~Banados, C.~Teitelboim, and J.~Zanelli, ``Black hole in three-dimensional
  spacetime,'' {\em Physical Review Letters} {\bf 69} (1992), no.~13, 1849.

\bibitem{Sheikh-Jabbari:2014nya}
M.~M. Sheikh-Jabbari and H.~Yavartanoo, ``{On quantization of AdS$_{3}$ gravity
  I: semi-classical analysis},'' {\em JHEP} {\bf 07} (2014) 104,
\href{http://www.arXiv.org/abs/1404.4472}{{\tt 1404.4472}}.

\bibitem{Sheikh-Jabbari:2016unm}
M.~M. Sheikh-Jabbari and H.~Yavartanoo, ``{On 3d bulk geometry of Virasoro
  coadjoint orbits: orbit invariant charges and Virasoro hair on locally
  AdS$_3$ geometries},'' {\em Eur. Phys. J.} {\bf C76} (2016), no.~9, 493,
\href{http://www.arXiv.org/abs/1603.05272}{{\tt 1603.05272}}.

\bibitem{Banados:1992gq}
M.~Ba\~nados, M.~Henneaux, C.~Teitelboim, and J.~Zanelli, ``Geometry of the
  (2+1) black hole,'' {\em Phys. Rev.} {\bf D48} (1993) 1506--1525,
\href{http://www.arXiv.org/abs/gr-qc/9302012}{{\tt gr-qc/9302012}}.

\bibitem{Banados:1998gg}
M.~Ba\~nados, ``{Three-dimensional quantum geometry and black holes},''
\href{http://www.arXiv.org/abs/hep-th/9901148}{{\tt hep-th/9901148}}.

\bibitem{brown1986poisson}
J.~D. Brown and M.~Henneaux, ``On the poisson brackets of differentiable
  generators in classical field theory,'' {\em Journal of mathematical physics}
  {\bf 27} (1986), no.~2, 489--491.

\bibitem{Compere:2018aar}
G.~Compère and A.~Fiorucci, ``{Advanced Lectures on General Relativity},''
  {\em Lect. Notes Phys.} {\bf 952} (2019) 150,
\href{http://www.arXiv.org/abs/1801.07064}{{\tt 1801.07064}}.

\bibitem{Hajian:2015xlp}
K.~Hajian and M.~M. Sheikh-Jabbari, ``{Solution Phase Space and Conserved
  Charges: A General Formulation for Charges Associated with Exact
  Symmetries},'' {\em Phys. Rev.} {\bf D93} (2016), no.~4, 044074,
\href{http://www.arXiv.org/abs/1512.05584}{{\tt 1512.05584}}.

\bibitem{Compere:2016hzt}
G.~Comp{\`e}re and J.~Long, ``{Classical static final state of collapse with
  supertranslation memory},'' {\em Class. Quant. Grav.} {\bf 33} (2016),
  no.~19, 195001,
\href{http://www.arXiv.org/abs/1602.05197}{{\tt 1602.05197}}.

\bibitem{Compere:2016jwb}
G.~Comp{\`e}re and J.~Long, ``{Vacua of the gravitational field},''
\href{http://www.arXiv.org/abs/1601.04958}{{\tt 1601.04958}}.

\bibitem{Compere:2015bca}
G.~Comp{\`e}re, K.~Hajian, A.~Seraj, and M.~M. Sheikh-Jabbari, ``{Wiggling
  Throat of Extremal Black Holes},'' {\em JHEP} {\bf 10} (2015) 093,
  \href{http://www.arXiv.org/abs/1506.07181}{{\tt 1506.07181}}.
[JHEP10,093(2015)].

\bibitem{Compere:2015mza}
G.~Compère, K.~Hajian, A.~Seraj, and M.~M. Sheikh-Jabbari, ``{Extremal
  Rotating Black Holes in the Near-Horizon Limit: Phase Space and Symmetry
  Algebra},'' {\em Phys. Lett.} {\bf B749} (2015) 443--447,
  \href{http://www.arXiv.org/abs/1503.07861}{{\tt 1503.07861}}.
[Phys. Lett.B749,443(2015)].

\end{thebibliography}

\providecommand{\href}[2]{#2}\begingroup\raggedright\endgroup

\end{document}